# Reshaping Cellular Networks for the Sky: Major Factors and Feasibility

Mohammad Mahdi Azari[1], Fernando Rosas[2,3] and Sofie Pollin[1]
[1] Department of Electrical Engineering, KU Leuven, Belgium
[2] Centre of Complexity Science and Department of Mathematics, Imperial College London, UK
[3] Department of Electrical and Electronic Engineering, Imperial College London, UK

***Abstract*—This paper studies the feasibility of supporting drone operations using existent cellular infrastructure. We propose an analytical framework that includes the effects of base station (BS) height and antenna radiation pattern, drone antenna directivity and various propagation environments. With this framework, we derive an exact expression for the coverage probability of ground and drone users through a practical cell association strategy. Our results show that a carefully designed network can control the radiated interference that is received by the drones, and therefore guarantees a satisfactory quality of service. Moreover, as the network density grows the increasing level of interference can be partially managed by lowering the drone flying altitude. However, even at optimal conditions the drone coverage performance converges to zero considerably fast, suggesting that ultra-dense networks might be poor candidates for serving aerial users.**

## I. Introduction

### A. Motivation

A widespread use of unmanned aerial vehicles (UAVs) in novel civil applications is currently being enabled by recent advances in the design of reliable and cost-effective drone technology. Scenarios include surveillance and monitoring, search and rescue operations, remote sensing, product delivery and many others [1]. All these applications depend critically on having reliable communications between drones and ground stations, particularly when drones require a beyond visual line-of-sight (LoS) tetherless connectivity.

The wireless connectivity for drones operation serve two main purposes: command/control and data communication. The former enables drone traffic management in remote missions, requiring *high coverage, low latency and continuous connectivity* [2]. In contrast, many drone use cases require *high speed data rates* for enabling real time delivery of telemetry data or high-resolution photographies.

In order to satisfy the aforementioned requirements, and in turn enable reliable wireless connectivity for drone applications, the cellular and Long-Term Evolution (LTE) technology seems to be an adequate choice [3], [4]. However, there are several challenges that must be addressed before deploying the LTE cellular technology in drones at large scale. In effect, cellular networks have been designed and optimized for serving ground users e.g. by using appropriate base stations (BSs) downtilt angles, which might result in significant antenna gain reductions for aerial users. Considering this, one might ask: are such networks capable of providing coverage on the sky? Moreover, is it possible for a cleverly designed drones network to take advantage of existing ground infrastructure without modifying the network technology and BS antenna configuration?

### B. Related Works

A few recent reports have addressed the above-mentioned questions via field trials [3], [4]. However, these works lack of modeling efforts, and therefore their results cannot explore the impact of various key parameters in the search of guidelines to support future developments of this technology. A first attempt to provide a theoretical perspective to these issues can be found in [5], where the coverage performance of a *cellular-connected drone* is studied when the drone connects to the closest BS. However, a more realistic assumption is that drones choose to associate with the BS from which it receives a strongest signal. Interestingly, it is not uncommon for UAVs to receive a stronger power from a BS that is not geographically the closest one. The first reason for this is that, depending on the downtilt angle of the BSs, a farther BS via its mainlobe may provide an stronger signal than a closer BS via its sidelobe. Secondly, the closest BS can be blocked by some obstacles and hence the received signal power might significantly drop.

Drone as aerial BS, however, has been studied in the majority of recent reports. In [6] we analyzed the downlink coverage performance of Poisson distributed drone BSs that provide wireless access for urban ground users. The results reveal that an altitude-dependent optimization of drones antenna beamwidth and density considerably mitigates interference and leads to significant improvement in the network performance. Moreover, [6] shows that a ground user in a denser urban environment can benefit from less interference due to the presence of more obstacles. In [7], [8] the aerial BS location is optimized to increase the coverage region and to lower the required transmission power through a novel proposed channel model which includes elevation angle-dependent path loss exponent and fading parameter. Furthermore, the optimal deployment of multiple aerial BSs for maximum total coverage region is analyzed in [9]. In addition, the aerial BSs backhaul has been addressed through several recent reports [10], [11].

### C. Contribution, Our Approach and Paper Structure

Our approach is to leverage available knowledge on ground cellular network analysis [12], [13] and generalize results for elevated users. To this end, some network properties that are often neglected in the stochastic analysis of ground-to-ground cellular networks play major roles. For instance, the impact of 3D features of the BSs' antenna patterns is particularly significant for aerial users, and should be taken into account.

The present paper extends and refines our previous work in [5] by considering a more practical cell association scenario towards to strongest BS, and also addressing the impact of the directivity of the drone's antenna and the thermal noise. Our results show different trends for ground and aerial users in terms of BS height. Concretely, we show that there exists an optimum BS height from a ground user perspective, which decreases as the BS density grows. Moreover, due to the different propagation conditions, elevating the BS could be disadvantageous for drone users.

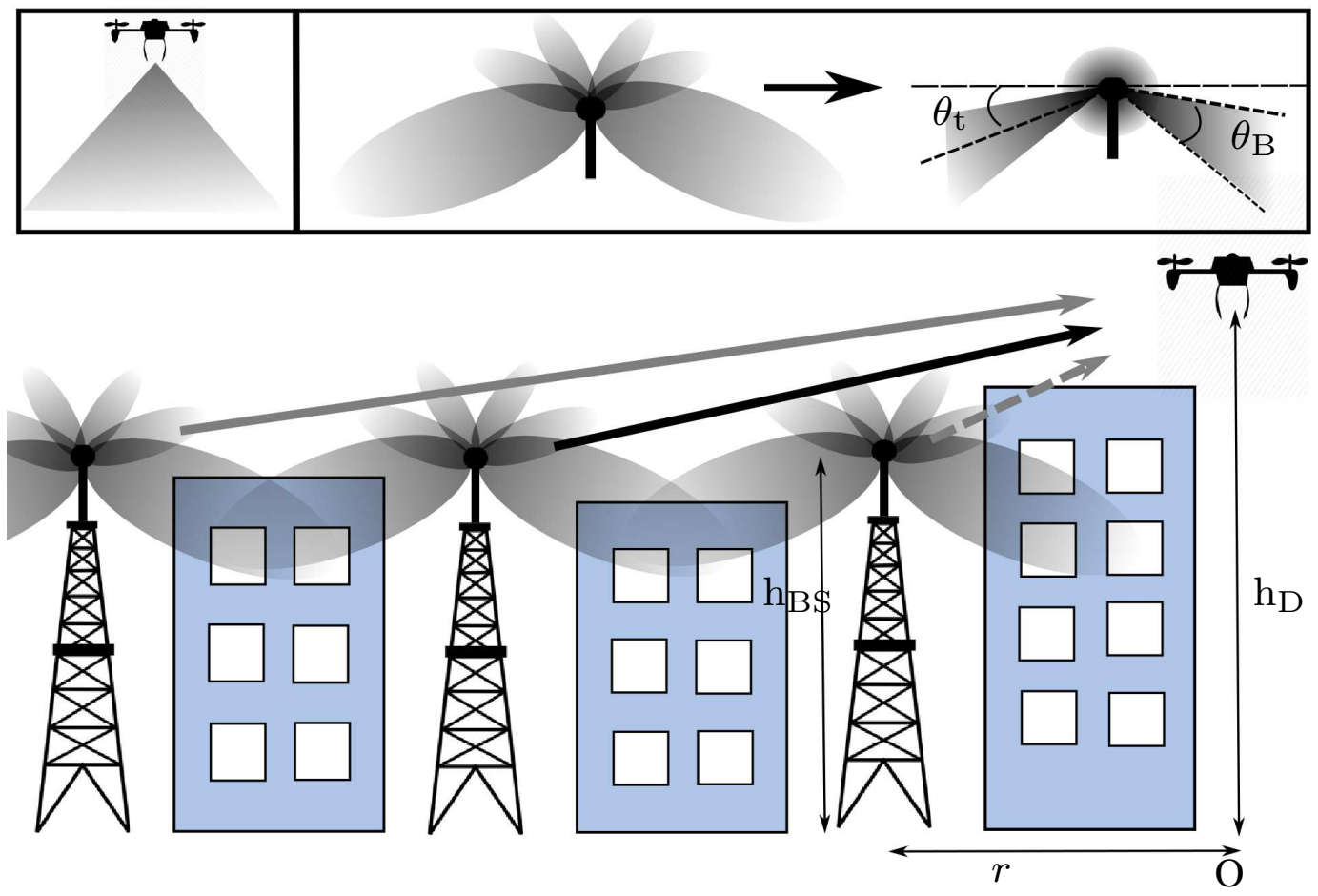

Fig. 1. *Below:* considered scenario, where the serving BS (represented by a bold arrow) can be interfered by other BSs located closer or further from the node. *Above:* models for the (simplified) directional antenna used by the drone UE and BS.

Our results show that drone users are LoS and interference limited in cellular networks, while NLoS links and noise effects are negligible. This motivates further explorations about LoS aware protocols. Furthermore, by studying the impact of random BS heights and downtilt angles distributions, it is shown that the corresponding coverage performance is well approximated by a network with fixed values equal to their averages. We also investigate the impact of antenna directionality at the drone. Our results illustrate how the antenna of the aerial user can be optimized at different altitudes in order to mitigate the effect of aggregate interference.

Finally, this work explores the impact of network densification, which shows different trends for ground and aerial users. Interestingly, as the network density increases, the flying altitude should be lowered to benefit from interference blocking by the obstacles.

The rest of this paper is organized as follows. First, the network model is introduced in Section II. The analysis of the coverage probability is presented in Section III, which is then verified by numerical evaluations and simulations in Section IV. Section IV also presents an extensive analysis of the impact of various network parameters. Finally, our main conclusions are summarized in Section V.

## II. Network Model

This section introduces the network architecture in Section II-A, the channel model in Section II-B, the BS association method and blockages modeling in urban areas in Sections II-C and II-D, respectively.

### A. System Architecture

We consider a ground cellular network serving ground and drone user equipments (UEs), i.e. G-UEs and D-UEs. The cellular network is formed by base stations (BSs) randomly distributed according to a homogeneous Poisson point process (HPPP) $\Phi$ of a fixed density $\lambda$ $\mathrm{BSs/Km^2}$. A BS is characterized by its height $\mathrm{h_{BS}}$, ground distance $r$ to the origin O, and its antenna radiation pattern. For analytical tractability, we adopt a sectored model to approximate the actual pattern of the antenna as is illustrated in Figure 1. To this end, we assume that the antenna radiation pattern is omnidirectional in the horizontal plane and vertically directional, with the beamwidth and downtitlt angle denoted by $\theta_{\mathrm{B}}$ and $\theta_{\mathrm{t}}$ respectively. The total power gains provided by the mainlobe and sidelobe of the BS antenna are denoted by $\mathrm{g_m}$ and $\mathrm{g_s}$, respectively.

A typical G-UE employs omnidirectional antenna and is located in the origin whereas a D-UE is placed $\mathrm{h_D}$ meter above the origin. In this work we consider the case where a D-UE employs a directional antenna pointing directly downwards bellow the drone and hence the antenna pattern has a beamwidth of $\varphi_{\mathrm{B}}$ (see Figure 1). The drone antenna gain can be approximated by $\mathrm{g_D} = 29000/\varphi_{\mathrm{B}}^2$ [14] within the main lobe and zero outside of the main lobe. Therefore, the drone receives the signal only from BSs within a ground circular region of radius $\Delta_{\mathrm{h}} \cdot \tan(\varphi_{\mathrm{B}}/2)$ where $\Delta_{\mathrm{h}} = \mathrm{h_D} - \mathrm{h_{BS}}$ and $\mathrm{h_D} > \mathrm{h_{BS}}$. Finally, the communication link distance $d$ between a BS at the ground distance $r$ and a UE can be obtained as $d = \sqrt{r^2 + \Delta_{\mathrm{h}}^2}$ where by $\mathrm{h_D} = 0$ the corresponding distance for a G-UE is obtained. In the following we use $\mathrm{r_{max}} = \Delta_{\mathrm{h}} \cdot \tan(\varphi_{\mathrm{B}}/2)$ for a D-UE and $\mathrm{r_{max}} = \infty$ for a G-UE. The distance $\mathrm{r_{max}}$ represents the radius of a circular region centered at O which contains the serving and interfering BSs.

### B. Channel Model

To model the communication channel, we consider LoS and non-LoS (NLoS) links separately along with their probabilities of occurrence. The path loss for each link can be expressed as

$$\zeta_v(r) = \mathrm{A}_v d^{-\alpha_v} = \mathrm{A}_v \left(r^2 + \Delta_{\mathrm{h}}^2\right)^{-\alpha_v/2}; \ v \in \{\mathrm{L}, \mathrm{N}\}, \quad (1)$$

where $v$ represents the type of link which is either LoS or NLoS, $\alpha_v$ is the path loss exponent corresponding to the link of type $v$, and $\mathrm{A}_v$ is a constant parameter representing the path loss at the reference distance $d = 1\mathrm{m}$, which differs for each LoS and NLoS component.

Furthermore, we consider independent small scale fading, whose instantaneous power is captured by the random variable $\Omega_v$. Without loss of generality we assume that $\mathbb{E}\{\Omega_v\} = 1$. In order to have the flexibility to study various propagation environments, we adopt the well-known Nakagami-m model [15]. Correspondingly, the cumulative distribution function (CDF) of $\Omega_v$ is given by

$$F_{\Omega_v}(\omega) \triangleq \mathbb{P}[\Omega_v < \omega] = 1 - \sum_{k=0}^{m_v - 1} \frac{(m_v \omega)^k}{k!} \exp(-m_v \omega), \quad (2)$$

where $m_v$ is the fading parameter which is assumed to be a positive integer for the sake of analytical tractability. Above, a larger $m_v$ corresponds to a lighter fading and hence an LoS link adopts a larger value than an NLoS link, i.e. $m_{\mathrm{L}} > m_{\mathrm{N}}$.

The received power at a UE from an LoS and NLoS BS[1] at the distance $r$ can be expressed as

$$P_{\rm rx}(r) = {\rm P_{tx}}\, {\rm g}(r)\zeta_v(r)\Omega_v, \tag{3}$$

where $\rm P_{tx}$ is the BS transmitted power, ${\rm g}(r)$ represents the total antenna directivity gain between the BS and the user, which can be written as

$${\rm g}(r) = \begin{cases} {\rm g_m}\cdot{\rm g_{UE}} & ; \text{ if } r\in\mathcal{S}_{\rm BS}\ \&\ r\in\mathcal{S}_{\rm UE} \\ {\rm g_s}\cdot{\rm g_{UE}} & ; \text{ if } r\notin\mathcal{S}_{\rm BS}\ \&\ r\in\mathcal{S}_{\rm UE} \\ 0 & ; \text{ otherwise} \end{cases} \tag{4}$$

where $\rm g_{UE}$ is equal to $29000/\varphi_{\rm B}^2$ for a D-UE and unit for a G-UE[2], $\mathcal{S}_{\rm BS}$ is formed by all the distances $r$ satisfying ${\rm h_{BS}} - r\tan(\theta_{\rm t}+\theta_{\rm B}/2) < {\rm h_D} < {\rm h_{BS}} - r\tan(\theta_{\rm t}-\theta_{\rm B}/2)$, the set $\mathcal{S}_{\rm UE}$ contains the distances $r$ with $r < {\rm r_{max}}$ where ${\rm r_{max}} = \Delta_{\rm h}\tan(\varphi_{\rm B}/2)$ for a D-UE and ${\rm r_{max}}=\infty$ for a G-UE.

### C. User Association and Link SINR

In this paper we consider a practical user association strategy in which a user is connected to the BS that provides the strongest signal. In other words, assuming the same transmitted power $\rm P_{tx}$ for all the BSs, the user is associated to the BS with $\max\{{\rm g}(r)\cdot\zeta_v(r)\}$. Due to the random location of the BSs, the serving BS ground distance $R_{\rm S}$ to the UE is random which can be expressed as

$$R_{\rm S} = \arg\max_{r\in\Phi}\ {\rm g}(r)\cdot\zeta_v(r). \tag{5}$$

We note that the serving BS can be either LoS or NLoS, and may serve the UE via its mainlobe or sidelobe. Moreover, due to the effect of blockages and also the antenna gain variation of BSs at different distances with respect to the user, $R_{\rm S}$ is not necessarily the closest BS. This fact is numerically evaluated in Section IV.

The communication link between a user and its serving BS is interfered by all the other BSs. Accordingly, the aggregate interference can be written as

$$I = \sum_{r\in\Phi\setminus R_{\rm S}} P_{\rm rx}(r). \tag{6}$$

Assuming that $\rm N_0$ is the noise power, the instantaneous signal-to-interference-plus-noise ratio (SINR) can be stated as

$$\mathsf{SINR} = \frac{P_{\rm rx}(R_{\rm S})}{I+{\rm N_0}}. \tag{7}$$

### D. Blockages Modeling and LoS Probability

In order to obtain an expression for the probability of LoS between a transmitter and a receiver at different heights, an urban area is modeled as a set of buildings located in a square grid in [16]. The 3D blockages are then characterized by the fraction of the total land area occupied by the buildings denoted by $a$, the mean number of buildings per ${\rm km}^2$ denoted by $b$, and the buildings height which is modeled by a Rayleigh probability density function (PDF) with ascale parameter $c$. Using this model, the proposed expression for the LoS probability between a BS of the height $\rm h_{BS}$ and a UE at an altitude $\rm h_D$, which are $r$ meters away, can be expressed as

$$\mathcal{P}_{\rm L}(r) = \prod_{n=0}^{m}\left[1-\exp\left(-\frac{\left[{\rm h_{BS}} - \frac{(n+0.5)({\rm h_{BS}}-{\rm h_D})}{m+1}\right]^2}{2c^2}\right)\right], \tag{8}$$

where $m = \lfloor\frac{r\sqrt{ab}}{1000}-1\rfloor$. Moreover, the probability of NLoS is $\mathcal{P}_{\rm N}(r) = 1-\mathcal{P}_{\rm L}(r)$. We note that $\mathcal{P}_{\rm L}(r)$ in (8) is a decreasing step function of $r$ and an increasing function of $\rm h_D$. The density of the environment[3] can be determined by varying the set of $(a,b,c)$.

Considering different communication links, the LoS probabilities are assumed to be independent meaning that we ignore the possible correlations of the blockage effects on the different links to ease the exact analysis. Therefore, the LoS BS process $\Phi_{\rm L}$ and NLoS BS process $\Phi_{\rm N}$ form two independent non-homogeneous PPP of density $\lambda_{\rm L}(r) = \lambda\mathcal{P}_{\rm L}(r)$ and $\lambda_{\rm N}(r) = \lambda\mathcal{P}_{\rm N}(r)$ respectively. Accordingly, $\Phi = \Phi_{\rm L}\cup\Phi_{\rm N}$ and $\lambda = \lambda_{\rm L}+\lambda_{\rm N}$.

## III. Performance Analysis

In this section we evaluate the performance of the downlink communication link for both G-UE and D-UE in terms of coverage probability $\mathcal{P}_{\rm cov}$ which is defined as

$$\mathcal{P}_{\rm cov} \triangleq \mathbb{P}[\mathsf{SINR} > {\rm T}]. \tag{9}$$

In the following we derive the coverage probability.

**Theorem 1.** *The exact coverage probability can be obtained as*

$$\mathcal{P}_{\rm cov} = \sum_{v\in\{{\rm L,N}\}}\int_0^{\rm r_{max}} \mathcal{P}^v_{{\rm cov}|R_{\rm S}}\ f^v_{R_{\rm S}}({\rm r_S})\ {\rm dr_S}. \tag{10}$$

*Above,* $f^v_{R_{\rm S}}({\rm r_S})$ *is the probability density function (PDF) of the serving BS ground distance, i.e.* $R_{\rm S}$*, which can be obtained as*

$$f^v_{R_{\rm S}}({\rm r_S}) = 2\pi\lambda_v({\rm r_S})\,{\rm r_S}\cdot\prod_{\xi\in\{{\rm L,N}\}} e^{-2\pi\int_{\mathcal{A}^v_{{\rm no}\xi}({\rm r_S})}\lambda_\xi(r)\ r{\rm d}r} \tag{11}$$

*where* $v\in\{{\rm L,N}\}$*, the sets* $\mathcal{A}^v_{\rm noL}({\rm r_S})$ *and* $\mathcal{A}^v_{\rm noN}({\rm r_S})$ *are obtained in Appendix and contain the LoS and NLoS BSs distances* $r$*, respectively, that can provide stronger signals to the UE as compared to the serving BS of type* $v$ *(being LoS or NLoS) at the distance* $\rm r_S$*.*

*Moreover,* $\mathcal{P}^v_{{\rm cov}|R_{\rm S}}$ *is the conditional coverage probability, given the serving BS distance and its type* $v$*, which can be found as*

$$\mathcal{P}^v_{{\rm cov}|R_{\rm S}} = \sum_{k=0}^{m_v-1}(-1)^k q_k\cdot\frac{d^k}{ds_v^k}\mathcal{L}^v_{I|R_{\rm S}}(s_v);\ v\in\{{\rm L,N}\} \tag{12}$$

*where*

$$q_k = \frac{e^{-{\rm N_0}s_v}}{k!}\sum_{j=k}^{m_v-1}\frac{{\rm N_0}^{j-k}s_v^j}{(j-k)!}, \tag{13}$$

$$s_v = \frac{m_v{\rm T}}{{\rm P_{tx}}\,{\rm g}({\rm r_S})\,\zeta_v({\rm r_S})}, \tag{14}$$

[1]A BS is called NLoS (LoS) if and only if there is (no) blockage intersecting its communication link to the typical user.

[2]Please note that later on we show that the communication link for a drone is interference limited and hence the gain of its antenna can be assumed to be unit as well, since it has the same impact on the received signal and aggregate interference.

[3]The *environment* density refers to the size, height and number of buildings in the urban area, which is categorized as Suburban, Urban, Dense Urban and Highrise Urban in [16]. However a dense *network* refers to a large density of BSs $\lambda$.

*and $\mathcal{L}^{\upsilon}_{I|R_{\rm S}}(s_\upsilon)$ is the Laplace transform of the conditional aggregate interference $I|R_{\rm S}$ evaluated at $s_\upsilon$ for the serving BS of type $\upsilon$.*

*Finally, $\mathcal{L}^{\upsilon}_{I|R_{\rm S}}(\cdot)$ is obtained as*

$$\mathcal{L}^{\upsilon}_{I|R_{\rm S}}(s_\upsilon) = \prod_{\xi\in\{\rm L,N\}} e^{-2\pi\int_{\bar{\mathcal{A}}^{\upsilon}_{{\rm no}\xi}(\rm r_S)} \lambda_\xi(r)\,[1-\Upsilon_\xi(r,s_\upsilon)]\,r{\rm d}r} \tag{15}$$

*where $s_\upsilon$ is expressed in (14) and*

$$\Upsilon_\xi(r,s_\upsilon) = \left(\frac{m_\xi}{m_\xi + s_\upsilon {\rm P_{tx}}\,{\rm g}(r)\,\zeta_\xi(r)}\right)^{m_\xi}, \tag{16}$$

$$\bar{\mathcal{A}}^{\upsilon}_{{\rm no}\xi} = [0, {\rm r_{max}}]\backslash\mathcal{A}^{\upsilon}_{{\rm no}\xi}. \tag{17}$$

*Proof.* Please find Appendix. □

Due to the presence of LoS BSs for a drone-UE in the following theorem we simplify the above theorem for drone communication. This expression highlights the major network parameters needed to analyze a drone network. For instance, it shows that the NLoS BS channel model has negligible impact on the drone communication.

**Theorem 2.** *The impact of NLoS links and noise for a D-UE is negligible and hence its coverage probability can be approximated by eliminating several of derivations and integrals as*

$$\mathcal{P}_{\rm cov} \approx \int_0^{\rm r_{max}} \mathcal{P}^{\rm L}_{{\rm cov}|R_{\rm S}}\; f^{\rm L}_{\rm R_S}(\rm r_S)\;{\rm dr_S} \tag{18a}$$

*where*

$$f^{\rm L}_{\rm R_S}(\rm r_S) \approx 2\pi\lambda_{\rm L}(\rm r_S)\,\rm r_S \cdot e^{-2\pi\int_{\mathcal{A}^{\rm L}_{\rm noL}(\rm r_S)}\lambda_{\rm L}(r)\;\,r{\rm d}r}, \tag{18b}$$

$$\mathcal{P}^{\rm L}_{{\rm cov}|R_{\rm S}} \approx \sum_{k=0}^{m_{\rm L}-1}\frac{(-s_{\rm L})^k}{k!}\cdot\frac{d^k}{ds^k_{\rm L}}\mathcal{L}^{\rm L}_{I|R_{\rm S}}(s_{\rm L}), \tag{18c}$$

$$\mathcal{L}^{\rm L}_{I|R_{\rm S}}(s_{\rm L}) \approx e^{-2\pi\int_{\bar{\mathcal{A}}^{\rm L}_{\rm noL}(\rm r_S)}\lambda_{\rm L}(r)\,[1-\Upsilon_{\rm L}(r,s_{\rm L})]\,r{\rm d}r}. \tag{18d}$$

*Proof.* The impact of NLoS links can be eliminated by considering $\lambda_{\rm N} = 0$ in Theorem 1. Numerical results in Section IV show the accuracy of such simplification. □

## IV. Numerical and Simulation Results

In this section, the results obtained in Section III are validated using Monte-Carlo simulations. Furthermore, the effect of various system parameters are studied, which enable us to provide recommendations to enhance the quality of service for both the ground and drone UEs.

The default values for the network parameters are listed in Table I. Moreover, in the following in order to study the non-directionality impact of drone antenna as well we consider a wide beamwidth angle of drone antenna $\varphi_{\rm B} = 170^o$, unless mentioned otherwise.

**Altitude-Dependent LoS BSs and Aggregate Interference.** Figure 2a shows the number of LoS BSs and the distribution of the serving BS distance at different drone altitudes. Results show that the drone is able to find more LoS BSs at higher altitudes, and is served by a further BS. This fact *extends the coverage region of each BS in the air resulting in a different association pattern and hence new challenges in handover* which should be carefully addressed in future. It is worth noting that the mean serving BS distance $\bar{R}_{\rm S}$ does not necessarily increase with the altitude. In fact, the drone is likely to be served by further BS through its mainlobe rather than by the closer BS via its sidelobe depending on the downtilt angle of BSs. For a very large $\theta_{\rm t}$ with $\theta_{\rm t} - \theta_{\rm B}/2 \geq 0$, the BSs mainlobes are under the horizon and the closest BS on average is the strongest BS for the drone due to a shorter distance.

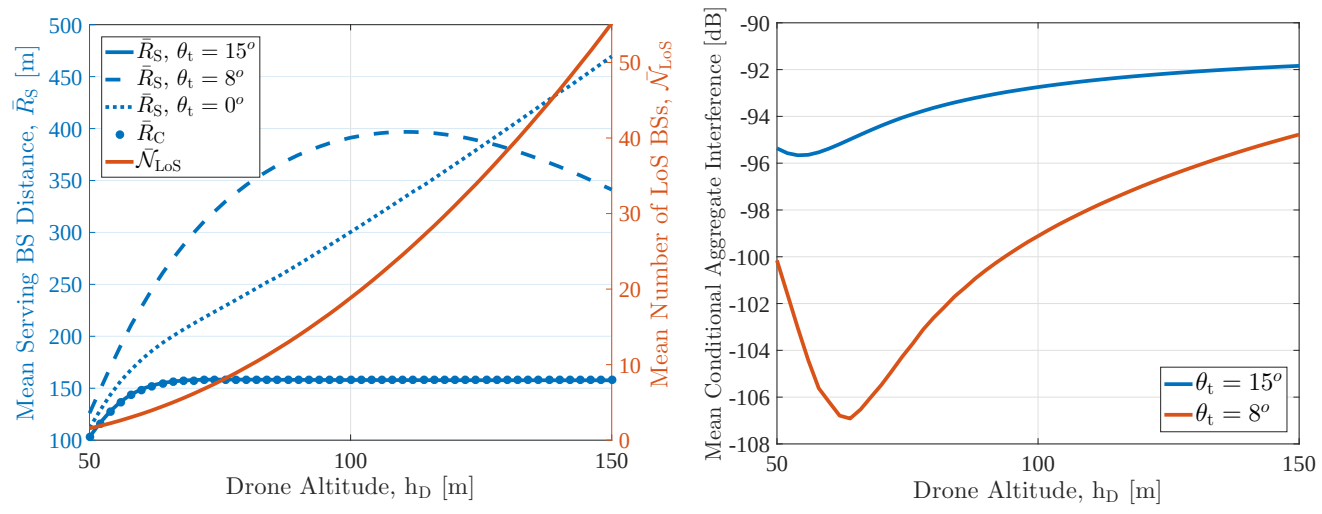

(a) (b)

Fig. 2. a) Depending on the downtilt angle $\theta_{\rm t}$ the drone connects to a further BS. In this figure $R_{\rm C}$ represents the ground distance to the closest BS. b) The interference behavior at different altitudes.

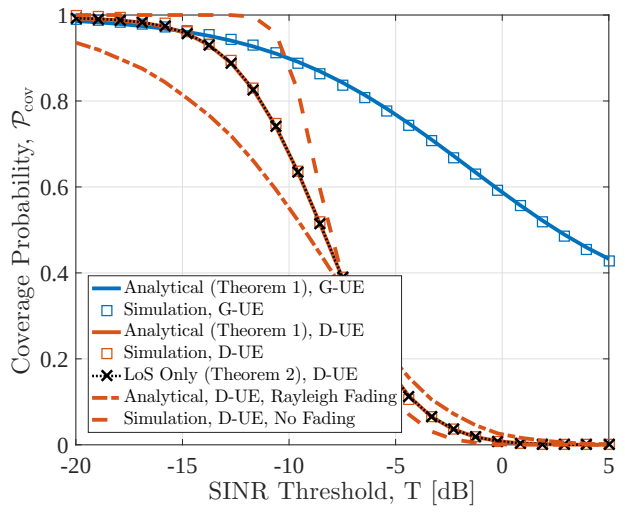

Fig. 3. The complementary CDF (CCDF) of SINR.

To get some insight into the behavior of the aggregate interference at different altitudes, Figure 2b illustrates the mean of conditional aggregate interference given the distance $\bar{R}_{\rm S}$. As can be seen, beyond a certain altitude, an increase in ${\rm h_D}$ will increase the aggregate interference power due to more interfering BSs that the drone can see.

TABLE I. Numerical result and simulation parameters.

| Parameter | Value |
|---|---|
| $(\alpha_{\rm L}, \alpha_{\rm N})$ | $(2.09\,,\,3.75)$ |
| $({\rm A_L}, {\rm A_N})$ | $(-41.1\,,\,-32.9)$ dB |
| $(m_{\rm L}, m_{\rm N})$ | $(1\,,\,3)$ |
| ${\rm P_{tx}}$ | $-6$ dB |
| T | 0.3 |
| $(a, b, c)$ | $(0.3\,,\,500\,,\,15)$ |
| $\lambda$ | 10 BSs/Km$^2$ |
| $(\theta_{\rm B}, \theta_{\rm t})$ | $(30^o\,,\,8^o)$ |
| $({\rm g_m}, {\rm g_s})$ | $(10\,,\,0.5)$ |
| ${\rm h_D}$ | 100m |
| ${\rm h_{BS}}$ | 30m |

**Validation of Theorem 1 and Accuracy of Theorem 2.** Figure 3 illustrates the complementary CDF (CCDF) of SINR for both ground and drone users. The Monte-Carlo simulations are done over $10^5$ network realizations. The figure shows that the analytical results are in a good conformity with the simulation. Moreover, the accuracy of the proposed approximation in Theorem 2 can be seen from the figure which confirms *the*

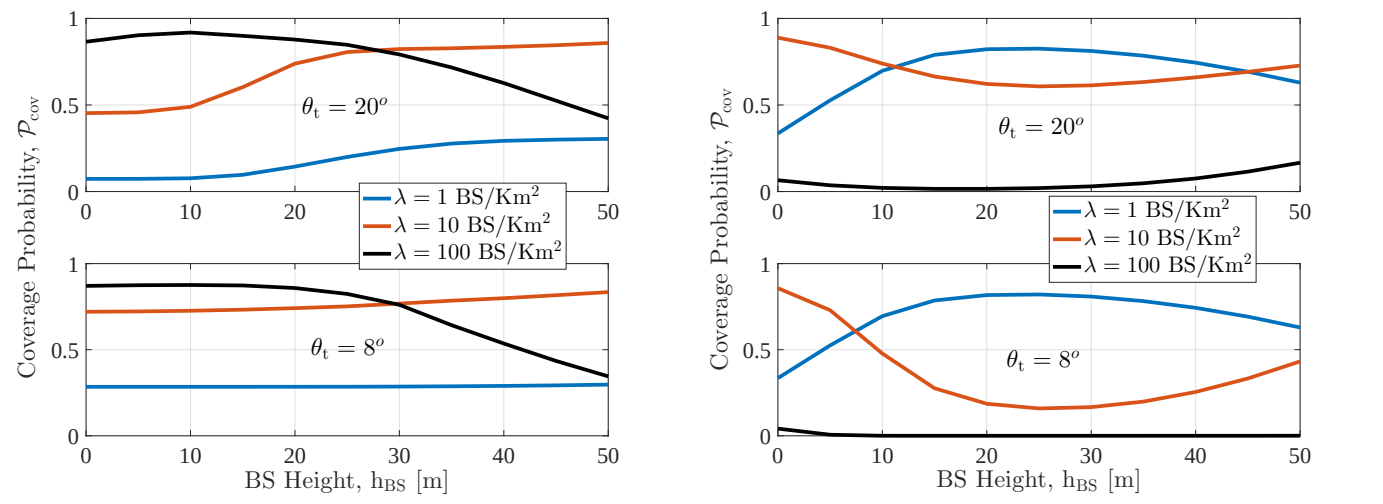

(a) G-UE. (b) D-UE at $\rm h_D$ =100m.

Fig. 4. Coverage probability versus different BS heights. The trends for ground and drone users are different.

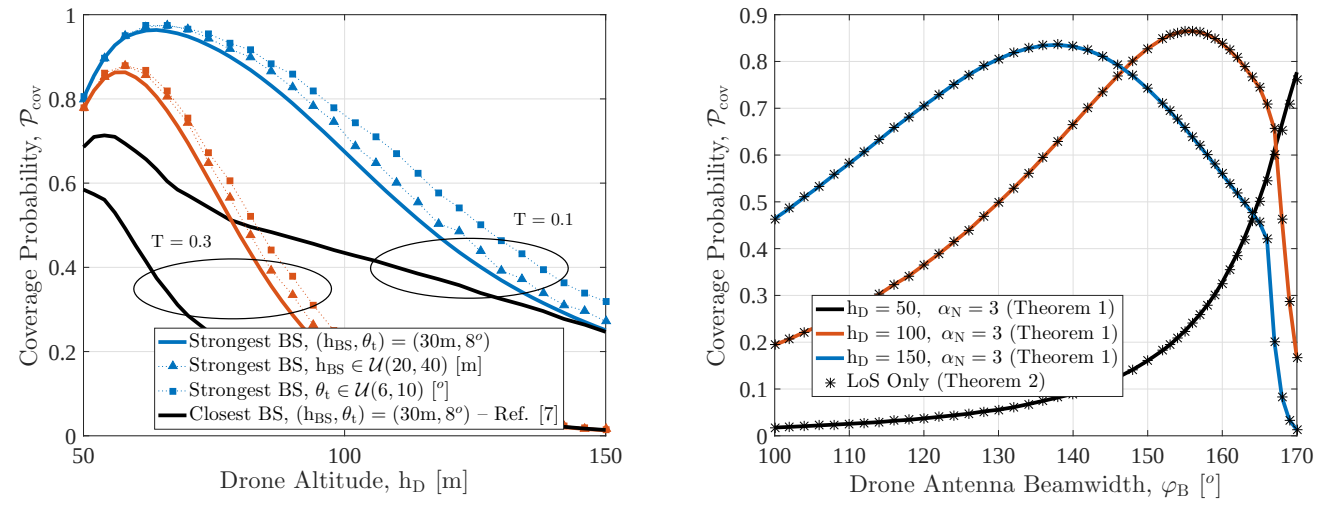

(a) (b)

Fig. 5. a) Random assumption for $\rm h_{BS}$ and $\theta_t$ has minor effect. b) The drone can be saved at different altitudes by optimizing its antenna beamwodth. The impact of noise and NLoS links for drone communication is negligible.

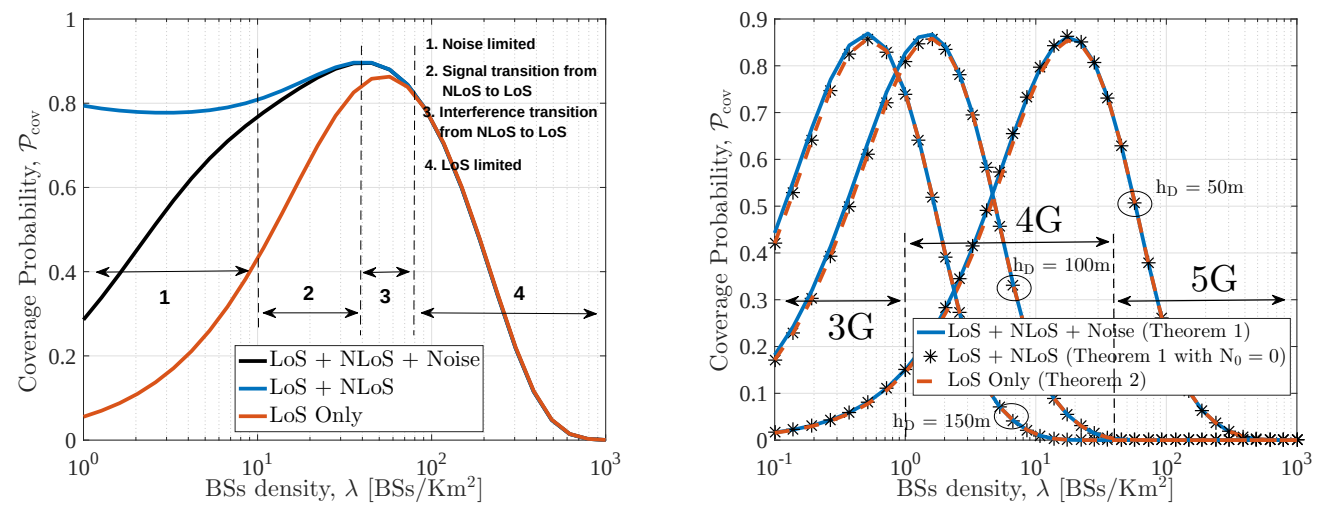

(a) (b)

Fig. 6. The effect of network densification for ground and drone users is detailed. The drone flying altitude should be lowered in denser networks to mitigate higher level of interference.

*impact of noise and NLoS links are negligible for a drone UE*. Moreover, the figure shows the effect of fading. In this figure Rayleigh fading corresponds to $m_L = m_N = 1$, whereas no fading case is simulated by adopting very large fading parameters which are $m_L = m_N = 100$. As can be seen *as the fading becomes lighter from Rayleigh to no fading, the SINR distribution starts concentrating.*

**BS Height.** Figure 4a shows that the BS height can play a major role in coverage such that *there exist an optimum height of BS for ground users and this optimum height decreases as the BS density increases.* In fact, for a low to medium BS density, an increase in BS height extends their mainlobe access region increasing the received signal power. However, for a large density of BSs this increases the interference level which in turn deteriorates the coverage performance. Comparing the curves corresponding to $\lambda = 10$ and $\lambda = 100$ show that *the height of BSs should be lowered for denser networks*.

*As compared to ground users the BS height adopts a different trend for drone users*. Figure 4b reveals that for a sparse network an increase in BS height can be beneficial up to some point due to the transition of serving BS from NLoS to LoS, however as the network goes dense, the increase in BS height is devastating due to the transition of interfering BSs from NLoS to LoS. Comparing with the ground users, the drone users coverage is better in sparse networks due to decreased interference. However as the network densifies, the performance of drone users drops. This implicitly means that *the impact of aggregate interference in the sky is significantly higher than the ground*.

**Drone Altitude.** On the one hand, there is a BS antenna gain reduction as the drone goes higher. On the other hand, the propagation condition will change from NLoS to LoS which is advantageous due to the increase in received signal and is disadvantageous since the aggregate interference also increases. These factors together result in an optimum altitude for maximum coverage as is shown in Figure 5a. Moreover, this figure shows that the coverage performance by considering the strongest BS as the serving cell is significantly higher than that is obtained in [5] by considering the closest BS.

**Distributed BS Height and Downtilt Angle.** Figure 5a illustrates the effect of uniformly distributed $\rm h_{BS}$ and $\theta_t$. As can be seen the variation in coverage probability is minor and the trend is precisely followed under the assumption of fixed values for $\rm h_{BS}$ and $\theta_t$ (equal to their means). In other words, *the impact of random assumptions for the BS height and downtilt angle is minor and can be neglected for the sake of analytical tractability.*

**Drone Directivity.** Figure 5b shows how the drone user can be efficiently equipped with a directional antenna for maximum coverage depending on the altitude of operation. In fact, as the drone beamwidth becomes wider there are more candidate BSs which results in a stronger serving BS, however a wider beamwidth leads to more interfering BSs within the mainlobe of the drone. These two opposite effects are balanced in an optimum beamwidth illustrated in the figure. Furthermore, *depending on $\varphi_B$ the drone operation altitude can be appropriately adjusted for maximum coverage.* For instance, as can be seen from the figure, at $\varphi_B = 160^0$ the drone operates safer in $\rm h_D = 100m$ compared to $\rm h_D = 50, 150m$. Furthermore, the optimum $\varphi_B$ decreases in higher altitudes to exclude more interfering BSs. Figure 5b also reveals that *even NLoS links with small corresponding path loss exponent do not affect the drone communication*.

**Network Densification.** *As the network densifies, the coverage probability for elevated users converges considerably faster to zero as compared to the ground user*, which is illustrated in Figure 6. This is due to the fact that the number of LoS BSs seen by the drones are significantly larger than ground users and the transition of NLoS interfering BSs to LoS occurs in lower $\lambda$s. The figure also show that *although the network for a D-UE is LoS limited, depending on $\lambda$ NLoS links play a major role for a G-UE*. Moreover, the noise can not be ignored for a G-UE as opposed to that of a D-UE. The figure, moreover, show that *the drone flying altitude should be lowered as the network goes dense* to benefit from interference blocking by the obstacles.

## V. Conclusion

The feasibility of using cellular networks for serving aerial users has been investigated. The most critical aspect for serving drones is to manage their extreme vulnerability to interference. Our findings suggest that interference can be successfully controlled by employing a carefully designed ground network in terms of BS height and downtilt angle, the drone antenna beamwidth and altitude. We showed that current cellular networks are capable of supporting drone-UEs. However, their integration to future ultra-dense networks will be challenging due to the high level of interference. Although some of these challenges can be addressed by choosing low flying altitudes and optimized drone antenna beamwidth, good integration eventually will require novel interference compensation techniques.

## Appendix A
## Proof of Theorem 1

In the following we use the notations:

$$[y]_x^+ \triangleq \max(x,y), \quad [y]_x^- \triangleq \min(x,y). \tag{19}$$

One can write

$$\mathcal{P}_{\rm cov} \triangleq \mathbb{P}[\mathsf{SINR} > \mathrm{T}] \tag{20}$$

$$= \sum_{v\in\{\mathrm{L,N}\}} \int_0^{\mathrm{r_{max}}} \mathcal{P}^v_{\mathrm{cov}|R_\mathrm{S}}\; f^v_{\mathrm{R_S}}(\mathrm{r_S})\, \mathrm{dr_S} \tag{21}$$

where

$$\mathcal{P}^{\mathrm{L}}_{\mathrm{cov}|R_\mathrm{S}} = \mathbb{P}[\mathsf{SINR} > \mathrm{T}|\, R_\mathrm{S} = \mathrm{r_S},\, \mathrm{LoS}], \tag{22}$$

$$\mathcal{P}^{\mathrm{N}}_{\mathrm{cov}|R_\mathrm{S}} = \mathbb{P}[\mathsf{SINR} > \mathrm{T}|\, R_\mathrm{S} = \mathrm{r_S},\, \mathrm{NLoS}] \tag{23}$$

are the conditional coverage probabilities when the distance $R_\mathrm{S}$ is given and the serving BS at the distance $R_\mathrm{S} = \mathrm{r_S}$ is LoS and NLoS respectively. Moreover, $f^{\mathrm{L}}_{\mathrm{R_S}}(\mathrm{r_S})$ and $f^{\mathrm{N}}_{\mathrm{R_S}}(\mathrm{r_S})$ represent the PDF of the distance $R_\mathrm{S}$ while the serving BS is LoS and NLoS respectively. Please note that no matter if the serving link is LoS or NLoS, they are interfered by both LoS and NLoS BSs.

Following the PPP properties, the function $f^{\mathrm{L}}_{\mathrm{R_S}}(\mathrm{r_S})$ can be written as

$$f^{\mathrm{L}}_{\mathrm{R_S}}(\mathrm{r_S}) = 2\pi\lambda_\mathrm{L}(\mathrm{r_S})\,\mathrm{r_S}\cdot \mathrm{P}^{\mathrm{L}}_{\mathrm{noL}}(\mathrm{r_S})\cdot \mathrm{P}^{\mathrm{L}}_{\mathrm{noN}}(\mathrm{r_S}), \tag{24}$$

where $2\pi\lambda_\mathrm{L}(\mathrm{r_S})\,\mathrm{r_S}$ is the unconditional PDF of having an LoS BS at the distance $\mathrm{r_S}$, $\mathrm{P}^{\mathrm{L}}_{\mathrm{noL}}(\mathrm{r_S})$ is the probability of having no LoS BS that provides stronger signal for the UE, and $\mathrm{P}^{\mathrm{L}}_{\mathrm{noN}}(\mathrm{r_S})$ is the probability of having no NLoS BS with better link. Assuming that $\mathcal{A}^{\mathrm{L}}_{\mathrm{noL}}(\mathrm{r_S})$ is formed by all the distances $r$ at which an LoS BS can provide a better link, $\mathrm{P}^{\mathrm{L}}_{\mathrm{noL}}(\mathrm{r_S})$ can be written as

$$\mathrm{P}^{\mathrm{L}}_{\mathrm{noL}}(\mathrm{r_S}) = e^{-2\pi\int_{\mathcal{A}^{\mathrm{L}}_{\mathrm{noL}}(\mathrm{r_S})} \lambda_\mathrm{L}(r)\; r dr}. \tag{25}$$

Similarly, if $\mathcal{A}^{\mathrm{L}}_{\mathrm{noN}}(\mathrm{r_S})$ is defined as the set of distances with stronger NLoS signal than the LoS signal at the distance $\mathrm{r_S}$, we have

$$\mathrm{P}^{\mathrm{L}}_{\mathrm{noN}}(\mathrm{r_S}) = e^{-2\pi\int_{\mathcal{A}^{\mathrm{L}}_{\mathrm{noN}}(\mathrm{r_S})} \lambda_\mathrm{N}(r)\; r dr}. \tag{26}$$

The sets $\mathcal{A}^{\mathrm{L}}_{\mathrm{noL}}(\mathrm{r_S})$ and $\mathcal{A}^{\mathrm{L}}_{\mathrm{noN}}(\mathrm{r_S})$ are dependent on the geometry of the network. In the following we derive the sets for the case of $\mathrm{h_D} > \mathrm{h_{BS}}$, however the similar approach can be employed to derive the sets for $\mathrm{h_D} \le \mathrm{h_{BS}}$. In order to obtain $\mathcal{A}^{\mathrm{L}}_{\mathrm{noL}}(\mathrm{r_S})$ and $\mathcal{A}^{\mathrm{L}}_{\mathrm{noN}}(\mathrm{r_S})$, we study two different cases separately where $\mathrm{r_0} = \Delta_\mathrm{h}\cdot\cot([\theta_\mathrm{B}/2-\theta_\mathrm{t}]_0^+)$:

1) $\mathrm{r_S} < \mathrm{r_0}$: In this case the LoS BS serves the UE from its sidelobe and hence one can write

$$\mathcal{A}^{\mathrm{L}}_{\mathrm{noL}}(\mathrm{r_S}) = [0,\mathrm{r_S}] \cup [\mathrm{r_0}, z_1], \tag{27}$$

where $[0,\mathrm{r_S}]$ contains the LoS BSs that can provide stronger signal by their sidelobes and $[\mathrm{r_0}, z_1]$ includes all the LoS BSs that can provide stronger signal by their mainlobes. The value of $z_1$ is obtained by solving the equation

$$\mathrm{P_{tx}g_m}\zeta_\mathrm{L}(r) = \mathrm{P_{tx}g_s}\zeta_\mathrm{L}(\mathrm{r_S}). \tag{28}$$

By taking the condition of $z_1 > \mathrm{r_0}$ into account the above equation yields $z_1$ expressed in Table II.

To calculate the set of $\mathcal{A}^{\mathrm{L}}_{\mathrm{noN}}(\mathrm{r_S})$ one can write

$$\mathcal{A}^{\mathrm{L}}_{\mathrm{noN}}(\mathrm{r_S}) = [0, z_2] \cup [\mathrm{r_0}, z_3], \tag{29}$$

where the first interval includes the NLoS BSs that can provide an stronger signal by their sidelobes and the second interval includes the NLoS BSs that can provide stronger signal by their mainlobes. The value of $z_2$ can be obtained from the equation

$$\mathrm{P_{tx}g_s}\zeta_\mathrm{N}(z_2) = \mathrm{P_{tx}g_s}\zeta_\mathrm{L}(\mathrm{r_S}). \tag{30}$$

Considering that $z_2 \ge 0$ is a real number, the above equation yields $z_2$ (see Table II). Now for $z_3$ we have

$$\mathrm{P_{tx}g_m}\zeta_\mathrm{N}(z_3) = \mathrm{P_{tx}g_s}\zeta_\mathrm{L}(\mathrm{r_S}). \tag{31}$$

Considering that $z_3 \ge \mathrm{r_0}$ is a real number, the above equation obtains $z_3$ that is given in Table II.

2) $\mathrm{r_S} \ge \mathrm{r_0}$: In this case the LoS BS at the distance $\mathrm{r_S}$ serves the UE through its mainlobe and hence we should have

$$\mathcal{A}^{\mathrm{L}}_{\mathrm{noL}}(\mathrm{r_S}) = [0, z_4] \cup [\mathrm{r_0}, \mathrm{r_S}] \tag{32}$$

$$\mathcal{A}^{\mathrm{L}}_{\mathrm{noN}}(\mathrm{r_S}) = [0, z_5] \cup [\mathrm{r_0}, z_6], \tag{33}$$

where the first and second intervals correspond to the BSs that can provide better link through their sidelobes and mainlobes respectively. Similar to the derivations above we can obtain the values $z_4$ to $z_6$ as is listed in table II.

Similarly, the PDF of a serving NLoS BS existed at the distance $\mathrm{r_S}$, i.e. $f^{\mathrm{N}}_{\mathrm{R_S}}(\mathrm{r_S})$, can be written as

$$f^{\mathrm{N}}_{\mathrm{R_S}}(\mathrm{r_S}) = 2\pi\lambda_\mathrm{N}(\mathrm{r_S})\,\mathrm{r_S}\cdot \mathrm{P}^{\mathrm{N}}_{\mathrm{noL}}(\mathrm{r_S})\cdot \mathrm{P}^{\mathrm{N}}_{\mathrm{noN}}(\mathrm{r_S}), \tag{34}$$

where $\mathrm{P}^{\mathrm{N}}_{\mathrm{noL}}(\mathrm{r_S})$ is the probability that there is no LoS BS providing better link and can be written as

$$\mathrm{P}^{\mathrm{N}}_{\mathrm{noL}}(\mathrm{r_S}) = e^{-2\pi\int_{\mathcal{A}^{\mathrm{N}}_{\mathrm{noL}}(\mathrm{r_S})} \lambda_\mathrm{L}(r)\; r dr}. \tag{35}$$

Moreover, the probability that there is no stronger NLoS signal from the other BSs is represented by $\mathrm{P}^{\mathrm{N}}_{\mathrm{noN}}(\mathrm{r_S})$ and is obtained as

$$\mathrm{P}^{\mathrm{N}}_{\mathrm{noN}}(\mathrm{r_S}) = e^{-2\pi\int_{\mathcal{A}^{\mathrm{N}}_{\mathrm{noN}}(\mathrm{r_S})} \lambda_\mathrm{N}(r)\; r dr}. \tag{36}$$

The sets $\mathcal{A}^{\mathrm{N}}_{\mathrm{noL}}(\mathrm{r_S})$ and $\mathcal{A}^{\mathrm{N}}_{\mathrm{noN}}(\mathrm{r_S})$ can be derived similar to $\mathcal{A}^{\mathrm{L}}_{\mathrm{noL}}(\mathrm{r_S})$ and $\mathcal{A}^{\mathrm{L}}_{\mathrm{noN}}(\mathrm{r_S})$ as

$$\begin{aligned}
\mathcal{A}^{\mathrm{N}}_{\mathrm{noL}}(\mathrm{r_S}) &= [0, z_7] \cup [\mathrm{r_0}, z_8];\ \text{for } \mathrm{r_S} < \mathrm{r_0}\\
\mathcal{A}^{\mathrm{N}}_{\mathrm{noN}}(\mathrm{r_S}) &= [0, \mathrm{r_S}] \cup [\mathrm{r_0}, z_9];\ \text{for } \mathrm{r_S} < \mathrm{r_0}\\
\mathcal{A}^{\mathrm{N}}_{\mathrm{noL}}(\mathrm{r_S}) &= [0, z_{10}] \cup [\mathrm{r_0}, z_{11}];\ \text{for } \mathrm{r_S} \ge \mathrm{r_0}\\
\mathcal{A}^{\mathrm{N}}_{\mathrm{noN}}(\mathrm{r_S}) &= [0, z_{12}] \cup [\mathrm{r_0}, \mathrm{r_S}];\ \text{for } \mathrm{r_S} \ge \mathrm{r_0}
\end{aligned} \tag{37}$$

where $z_i$s are listed in table II.

The following table summarizes the results for the sets $\mathcal{A}^{\upsilon}_{\mathrm{no}\xi}(\mathrm{r_S})$.

TABLE II. The values of $z_i$ for $i = 1, 2, 3, ..., 12$.

| $z_i$ |
|---|
| $z_1 = [\sqrt{\rho_1(\mathrm{r}_\mathrm{S}^2+\Delta_\mathrm{h}^2)-\Delta_\mathrm{h}^2}]_{\mathrm{r}_0}^{+};\ \rho_1=(\mathrm{g_m}/\mathrm{g_s})^{2/\alpha_\mathrm{L}}$ |
| $z_2 = \sqrt{[\rho_2(\mathrm{r}_\mathrm{S}^2+\Delta_\mathrm{h}^2)^{\alpha_\mathrm{L}/\alpha_\mathrm{N}}-\Delta_\mathrm{h}^2]_0^{+}};\ \rho_2=(\mathrm{A_N}/\mathrm{A_L})^{2/\alpha_\mathrm{N}}$ |
| $z_3 = \left[\sqrt{[\rho_3(\mathrm{r}_\mathrm{S}^2+\Delta_\mathrm{h}^2)^{\alpha_\mathrm{L}/\alpha_\mathrm{N}}-\Delta_\mathrm{h}^2]_0^{+}}\right]_{\mathrm{r}_0}^{+};\ \rho_3=\left(\frac{\mathrm{A_N g_m}}{\mathrm{A_L g_s}}\right)^{2/\alpha_\mathrm{N}}$ |
| $z_4 = \left[\sqrt{[\rho_4(\mathrm{r}_\mathrm{S}^2+\Delta_\mathrm{h}^2)-\Delta_\mathrm{h}^2]_0^{+}}\right]_{\mathrm{r}_0}^{-};\ \rho_4=1/\rho_1$ |
| $z_5 = \left[\sqrt{[\rho_5(\mathrm{r}_\mathrm{S}^2+\Delta_\mathrm{h}^2)^{\alpha_\mathrm{L}/\alpha_\mathrm{N}}-\Delta_\mathrm{h}^2]_0^{+}}\right]_{\mathrm{r}_0}^{-};\ \rho_5=\left(\frac{\mathrm{A_N g_s}}{\mathrm{A_L g_m}}\right)^{2/\alpha_\mathrm{N}}$ |
| $z_6 = \left[\mathrm{r}_0, \sqrt{[\rho_6(\mathrm{r}_\mathrm{S}^2+\Delta_\mathrm{h}^2)^{\alpha_\mathrm{L}/\alpha_\mathrm{N}}-\Delta_\mathrm{h}^2]_0^{+}}\right]_{\mathrm{r}_0}^{+};\ \rho_6=\rho_2$ |
| $z_7 = \left[\sqrt{\rho_7(\mathrm{r}_\mathrm{S}^2+\Delta_\mathrm{h}^2)^{\alpha_\mathrm{N}/\alpha_\mathrm{L}}-\Delta_\mathrm{h}^2}\right]_{\mathrm{r}_0}^{-};\ \rho_7=(\mathrm{A_L}/\mathrm{A_N})^{2/\alpha_\mathrm{L}}$ |
| $z_8 = \left[\sqrt{[\rho_8(\mathrm{r}_\mathrm{S}^2+\Delta_\mathrm{h}^2)^{\alpha_\mathrm{N}/\alpha_\mathrm{L}}-\Delta_\mathrm{h}^2]_0^{+}}\right]_{\mathrm{r}_0}^{+};\ \rho_8=\left(\frac{\mathrm{A_L g_m}}{\mathrm{A_N g_s}}\right)^{2/\alpha_\mathrm{L}}$ |
| $z_9 = \left[\sqrt{\rho_9(\mathrm{r}_\mathrm{S}^2+\Delta_\mathrm{h}^2)-\Delta_\mathrm{h}^2}\right]_{\mathrm{r}_0}^{+};\ \rho_9=(\mathrm{g_m}/\mathrm{g_s})^{2/\alpha_\mathrm{N}}$ |
| $z_{10} = \left[\sqrt{[\rho_{10}(\mathrm{r}_\mathrm{S}^2+\Delta_\mathrm{h}^2)^{\alpha_\mathrm{N}/\alpha_\mathrm{L}}-\Delta_\mathrm{h}^2]_0^{+}}\right]_{\mathrm{r}_0}^{-};\ \rho_{10}=\left(\frac{\mathrm{A_L g_s}}{\mathrm{A_N g_m}}\right)^{2/\alpha_\mathrm{L}}$ |
| $z_{11} = \left[\sqrt{\rho_{11}(\mathrm{r}_\mathrm{S}^2+\Delta_\mathrm{h}^2)^{\alpha_\mathrm{N}/\alpha_\mathrm{L}}-\Delta_\mathrm{h}^2}\right]_{\mathrm{r}_0}^{+};\ \rho_{11}=\rho_7$ |
| $z_{12} = \left[\sqrt{[\rho_{12}(\mathrm{r}_\mathrm{S}^2+\Delta_\mathrm{h}^2)-\Delta_\mathrm{h}^2]_0^{+}}\right]_{\mathrm{r}_0}^{-};\ \rho_{12}=1/\rho_9$ |

To obtain the conditional coverage probability $\mathcal{P}^{\upsilon}_{\mathrm{cov}|R_\mathrm{S}}$ One can write

$$\begin{aligned}
\mathcal{P}^{\upsilon}_{\mathrm{cov}|R_\mathrm{S}} &= \mathbb{P}\left[\frac{\mathrm{P_{tx}}\,\mathrm{g}(\mathrm{r_S})\,\zeta_\upsilon(\mathrm{r_S})\,\Omega_\upsilon}{\mathrm{N}_0+I} > \mathrm{T}\ \Big|\ R_\mathrm{S}=\mathrm{r_S}\right] \\
&= \mathbb{E}_I\left\{\mathbb{P}\left[\Omega_\upsilon > \frac{\mathrm{T}}{\mathrm{P_{tx}}\,\mathrm{g}(\mathrm{r_S})\,\zeta_\upsilon(\mathrm{r_S})}(\mathrm{N}_0+I)\right]\ \Big|\ R_\mathrm{S}=\mathrm{r_S}\right\} \\
&\overset{(a)}{=} \mathbb{E}_I\left\{\sum_{k=0}^{m_\upsilon-1}\frac{s_\upsilon^k}{k!}\,(\mathrm{N}_0+I)^k\exp[-s_\upsilon(\mathrm{N}_0+I)]\ \Big|\ R_\mathrm{S}=\mathrm{r_S}\right\} \\
&= \mathbb{E}_I\left\{\sum_{k=0}^{m_\upsilon-1}\frac{s_\upsilon^k}{k!}e^{-\mathrm{N}_0 s_\upsilon}\sum_{j=0}^{k}\binom{k}{j}\mathrm{N}_0^{k-j}I^j\exp[-s_\upsilon I]\ \Big|\ R_\mathrm{S}=\mathrm{r_S}\right\} \\
&= \sum_{k=0}^{m_\upsilon-1} q_k\cdot\mathbb{E}_I\left\{I^k\exp(-s_\upsilon I)\ \big|\ R_\mathrm{S}=\mathrm{r_S}\right\} \\
&= \sum_{k=0}^{m_\upsilon-1}(-1)^k q_k\cdot\frac{d^k}{ds_\upsilon^k}\mathcal{L}^{\upsilon}_{I|R_\mathrm{S}}(s_\upsilon),
\end{aligned} \tag{38}$$

where (a) follows from the gamma distribution of $\Omega_\upsilon$ with an integer parameter $m_\upsilon$ and $q_k$ and $s_\upsilon$ are expressed in (13) and (14) respectively.

To derive $\mathcal{L}_{I|R_\mathrm{S}}(s_\upsilon)$ one can write

$$\begin{aligned}
\mathcal{L}_{I|R_\mathrm{S}}(s_\upsilon) &= \mathbb{E}_I\{\exp(-s_\upsilon I)\ \mid\ R_\mathrm{S}=\mathrm{r_S}\} \\
&= \mathbb{E}_{\Phi,\Omega}\left\{\prod_{r\in\Phi\setminus\mathrm{r_S}}\exp[-s_\upsilon P_\mathrm{rx}(r)]\right\} \\
&= \mathbb{E}_{\Phi}\left\{\prod_{r\in\Phi\setminus\mathrm{r_S}}\mathbb{E}_\Omega\left\{\exp[-s_\upsilon P_\mathrm{rx}(r)]\right\}\right\}.
\end{aligned}$$

The above equation can be further processed as

$$\begin{aligned}
\mathcal{L}_{I|R_\mathrm{S}}(s_\upsilon) = \mathbb{E}_{\Phi_\mathrm{L}}&\left\{\prod_{r\in\Phi_\mathrm{L}\setminus\mathrm{r_S}}\mathbb{E}_\Omega\left\{\exp[-s_\upsilon P_\mathrm{rx}(r)]\right\}\right\} \\
\times\,\mathbb{E}_{\Phi_\mathrm{N}}&\left\{\prod_{r\in\Phi_\mathrm{N}\setminus\mathrm{r_S}}\mathbb{E}_\Omega\left\{\exp[-s_\upsilon P_\mathrm{rx}(r)]\right\}\right\} \\
\overset{(a)}{=}\ & e^{-2\pi\int_{\bar{\mathcal{A}}^{\upsilon}_{\mathrm{noL}}(\mathrm{r_S})}\lambda_\mathrm{L}(r)\,[1-\Upsilon_\mathrm{L}(r,s_\upsilon)]\,r\,dr} \\
\times\ & e^{-2\pi\int_{\bar{\mathcal{A}}^{\upsilon}_{\mathrm{noN}}(\mathrm{r_S})}\lambda_\mathrm{N}(r)\,[1-\Upsilon_\mathrm{N}(r,s_\upsilon)]\,r\,dr}
\end{aligned}$$

where (a) is obtained using the probability generating functional (PGFL) of PPP. Moreover, in the above equation $\bar{\mathcal{A}}^{\upsilon}_{\mathrm{noL}}$ and $\bar{\mathcal{A}}^{\upsilon}_{\mathrm{noN}}$ indicate the complementary of the sets $\mathcal{A}^{\upsilon}_{\mathrm{noL}}$ and $\mathcal{A}^{\upsilon}_{\mathrm{noN}}$ over the set $[0, \mathrm{r_{max}}]$ respectively.